\documentclass{ws-procs9x6}

\newcommand{\bea}   {\begin{eqnarray}}
\newcommand{\eea}   {\end{eqnarray}}
\begin{document}

% Write your title in CAPITALS in the next line [if necessary, use \\ to break lines]:
\title{On Chiral and Nonchiral $1D$ Supermultiplets}

% Your name(s) and address(es):
% Write names in CAPITALS; combine those with same addresses in one line\author{ *********** A. B. AUTHOR$^*$ and C. D. AUTHOR ********}

% In the address, use \\ to break lines, if necessary:
%\address{University Department, University Name,\\
%City, State ZIP/Zone, Country\\
%$^*$E-mail: ab\_author@university.com\\
%www.university\_name.edu}

\author{FRANCESCO TOPPAN}

\address{CBPF (TEO), Rua Dr. Xavier Sigaud 150,\\
cep 22290-180, Rio de Janeiro (RJ), Brazil\\
E-mail:  toppan@cbpf.br}

\begin{abstract}
% Replace next line with your abstract:
In this talk I discuss and clarify some issues concerning chiral and nonchiral properties of the
one-dimensional supermultiplets of the ${\cal N}$-Extended Supersymmetry. Quaternionic chirality can be defined for ${\cal N}=4,5,6,7, 8$. Octonionic chirality for ${\cal N}=8$ and beyond.
Inequivalent chiralities only arise when considering several copies of ${\cal N}=4$ or ${\cal N}=8$ supermultiplets.
\end{abstract}

% Enter your keywords:

\bodymatter

% Replace next two lines with the body of your paper (except Acknowledgements):
\section{Introduction}
The $1D$ ${\cal N}$-Extended Superalgebra, with ${\cal N}$ odd generators $Q_I$ ($I=1,2,\ldots, {\cal N}$)
and a single even generator $H$ satisfying the (anti)-commutation relations
\begin{eqnarray}\label{sqm}
\{ Q_I, Q_J\}&=& \delta_{IJ} H, \nonumber\\
\relax [H, Q_I]&=&0,
\end{eqnarray}
is the superalgebra underlying the Supersymmetric Quantum Mechanics \cite{wit}.
\par
In recent years the structure of its linear representations has been unveiled by a series of works  \cite{pt}--\cite{gkt}.\par
The linear representations under considerations (supermultiplets) contain a finite, equal number of bosonic and fermionic fields
depending on a single coordinate (the time). The operators $Q_I$ and $H$ act as differential operators. The
linear representations are characterized by a series of properties which, for sake of consistency, are reviewed in the appendix.\par
The {\em minimal} linear representations (also called {\em irreducible supermultiplets}) are given by
the minimal number $n_{min}$ of bosonic (fermionic) fields for a given value of ${\cal N}$.
The value $n_{min}$ is given \cite{pt} by the formula
\begin{eqnarray}\label{irrepdim}
{\cal N}&=& 8l+m,\nonumber\\
n_{min}&=& 2^{4l}G(m),
\end{eqnarray}
where $l=0,1,2,\ldots$ and $m=1,2,3,4,5,6,7,8$.\par
$G(m)$ appearing in (\ref{irrepdim}) is the Radon-Hurwitz function
\begin{eqnarray}&\label{radonhur}
\begin{tabular}{|c|cccccccc|}\hline
$m $&$1$&$2$&$3$& $4$&$5$&$6$&$7$&$8$\\ \hline
$G(m)  $&$1$&$2$&$4$& $4$&$8$&$8$&$8$&$8$\\ \hline
\end{tabular}&\nonumber\\
&&\end{eqnarray}
Non-minimal linear representations have been discussed in \cite{{krt},{dfghilm},{dfghilm2},{gkt}}.
\par
The construction of off-shell invariant actions (sigma-models) based on these representations has been given in \cite{{krt},{grt},{gkt}}. 
In this approach the superfields techniques that can be employed to recover invariant actions for the given supermultiplets \cite{abc} are no longer necessary.\par
In this talk I will address and clarify a specific issue.  In \cite{gh} (see also references therein) the  
${\cal N}=4$ minimal linear representations (given by the field content $(4,4)$, $(3,4,1)$, $(2,4,2)$ and $(1,4,3)$)  are given in two different forms, called ``chiral" and ``twisted chiral
supermultiplets". This double realization of the ${\cal N}=4$ representations can also be extended to the nonlinear realizations (the $(3,4,1)_{nl}$ and the $(2,4,2)_{nl}$ which are
recovered from the $(4,4)$ root supermultiplet via a specific construction based on the supersymmetrization of the first Hopf fibration, see \cite{fkt} and references therein).\par
For our purposes it is convenient to refer to this doubling as ``chirality". The two versions
of the supermultiplets will be conveniently denoted as ``chiral" and ``antichiral", respectively.  Chirality seems therefore encoded in all 
minimal (linear and non-linear) ${\cal N}=4$ representations.  On the other hand,
the $(4,4)$ root supermultiplet which generates all remaining ${\cal N}=4$ representations
(for the linear ones such as $(3,4,1)$, $(2,4,2)$ and $(1,4,3)$ through the dressing, see \cite{pt}
and the appendix), is in one-to-one correspondence with the (Weyl-type) realization of the $Cl(4,0)$ Clifford algebra \cite{pt}. This realization is unique. Therefore, a natural question to be asked is whether the notion of chirality is truly there. The answer, as we will see, is rather subtle. Since chirality (if there) is inherited from the properties of the root supermultiplets, for our purpose is
sufficient to address this problem for the root supermultiplets at a given ${\cal N}$. In particular, for ${\cal N}=4$, its $(4,4)$ root supermultiplet and, for ${\cal N}=8$, its $(8,8)$ root supermultiplet.    

\section{Quaternions, octonions and the ${\cal N}=4,8$ root supermultiplets: are they chiral?}

The ${\cal N}=4$ root supermultiplet of field content $(4,4)$ admits $4$ bosonic fields $x, x_i$
and $4$ fermionic fields $\psi, \psi_i$ ($i=1,2,3$). Its supertransformations are expressed in terms of the quaternionic structure constants ($\delta_{ij}$ and the totally antisymmetric tensor $\epsilon_{ijk}$). Without loss of generality we can explicitly realize them as
\bea\label{quat}&
\begin{array}{ll}
 Q_4 x=\psi,\quad\quad & Q_i x= \psi_i ,\\
Q_4 x_j =\psi_j,\quad \quad&Q_i x_j = -\delta_{ij} {\psi}+s\epsilon_{ijk} {\psi_k},\\
Q_4 \psi= {\dot x}, \quad\quad & Q_i \psi= -{\dot x_i},\\
Q_4  \psi_j={\dot x_j},\quad\quad &Q_i \psi_j= \delta_{ij} {\dot x}-s\epsilon_{ijk}  {\dot x_k}.
\end{array}&\nonumber\\
&&
\eea
A sign $s=\pm 1$ has been introduced. It corresponds to the convention of choosing the overall sign of the totally antisymmetric tensor $\epsilon_{ijk}$. It discriminates the chiral ($s=+1$) from the antichiral $(s=-1$) 
${\cal N}=4$ root supermultiplet. Due to its origin, we refer to this chirality as ``quaternionic chirality".\par
 A similar notion of chirality can be introduced for the ${\cal N}=8$  $(8,8)$ root supermultiplet which, without loss of generality, can be expressed \cite{krt} by replacing the quaternionic structure constant $\epsilon_{ijk}$  in (\ref{quat})  with the totally antisymmetric octonionic structure constants $C_{ijk}$, $i,j,k=1,2\ldots, 7$, such that 
$C_{123}=C_{147}=C_{165}=C_{246}=C_{257}=C_{354}=C_{367}=1$.\par
We have the following supertransformations acting on the $8$ bosonic fields $x,x_i$ and the $8$ fermionic fields $\psi, \psi_i$:
\bea\label{oct}&
\begin{array}{ll}
 Q_8 x=\psi,\quad\quad & Q_i x= \psi_i ,\\
Q_8 x_j =\psi_j,\quad \quad&Q_i x_j = -\delta_{ij} {\psi}+sC_{ijk} {\psi_k},\\
Q_8 \psi= {\dot x}, \quad\quad & Q_i \psi= -{\dot x_i},\\
Q_8  \psi_j={\dot x_j},\quad\quad &Q_i \psi_j= \delta_{ij} {\dot x}-sC_{ijk}  {\dot x_k}.
\end{array}&\nonumber\\
&&
\eea
The sign $s=\pm 1$ defines the ``octonionic chirality" of the ${\cal N}=8$ root supermultiplet.\par
It is easily realized that the chiral and antichiral supermultiplets are isomorphic and related by a $Z_2$ transformation. In the ${\cal N}=4$ case this is easily achieved by exchanging $x_2\leftrightarrow x_3$, $\psi_2\leftrightarrow\psi_3$ and by relabeling the supersymmetry transformations $Q_2\leftrightarrow Q_3$. It therefore looks that the chirality is an abusive notion which should be dismissed as useless. In the following we will prove that this is not quite so. It is certainly true that the notion of ``quaternionic chirality" turns out to be useless for the ${\cal N}=3$ $(4,4)$ root supermultiplet (it is obtained from (\ref{quat}) by disregarding the $Q_4$ supertransformation).
The difference w.r.t. the ${\cal N}=4$ case lies in the fact that the $Z_2$ isomorphism in this case can be imposed {\em without} relabeling the supertransformations (it is sufficient,
e.g., to map $\psi\mapsto -\psi$, $\psi_i\mapsto-\psi_i$, while leaving $x, x_i$ unchanged).
The relabeling of the supertransformations, which  is essential to implement the $Z_2$ isomorphism for the ${\cal N}=4$ quaternionic chirality (and the ${\cal N}=8$ octonionic chirality), makes all the difference w.r.t. the ${\cal N}=3$ root supermultiplet case. 

\section{Irreducible ${\cal N}=5,6,7$ supermultiplets are nonchiral, reducible supermultiplets are chiral}

The ${\cal N}=5,6,7,8$ root supermultiplets have field content $(8,8)$.  They all admit a
 decomposition into two minimal ${\cal N}=4$ supermultiplets obtained by suitably picking $4$ supertransformations out of the ${\cal N}$ original ones. For ${\cal N}=8$ we have $\left(\begin{array}{c} 8\\4\end{array}\right)=70$ inequivalent choices of $4$ supertransformations. $14$ of such choices
produce two minimal ${\cal N}=4$ supermultiplets (in the remaining $56$ cases we end up with a non-minimal, reducible but indecomposable, ${\cal N}=4$ representation \cite{gkt}). This number can be understood as follows: one can pick any $3$ supertransformations (necessarily ending up
with two ${\cal N}=3$ root supermultiplets). Then one is left with $5$ possible choices for the fourth
supertransformation. In $1$ case ($\frac{1}{5}$ of the total) we get
two separate minimal ${\cal N}=4$ supermultiplets. In the $4$ remaining cases we end up instead with
an indecomposable non-minimal ${\cal N}=4$ supermultiplet.\par
The ${\cal N}=8$ supertransformations in (\ref{oct}) can be associated with the octonions:
$Q_8$ with the octonionic
identity and the $Q_i$'s with the $7$ imaginary octonions. With this identification, the
$14$ combinations which produce a decomposition into two separate ${\cal N}=4$ supermultiplets are obtained as follows: $7$ by picking up $Q_8$ and $3$ supertransformations
lying in one of the $7$ lines of the Fano plane. The remaining $7$ by picking up $4$ supertransformations in the Fano plane which are complementary to one of the $7$ lines.\par
Incidentally, this counting proves that for ${\cal N}=5$ (and, {\em a fortiori}, ${\cal N}=6,7$)
one can always find a decomposition into two minimal ${\cal N}=4$ supermultiplets.
Indeed,  ${\cal N}=5$ supertransformations can be obtained in two ways: either, {\em case a},
with $5$ supertransformations lying on the Fano plane or, {\em case b}, $Q_8$ and $4$ supertransformations lying on the Fano plane. In case $a$, $4$ of the $5$ supertransformations are necessarily complementary to a line; in case $b$ we have two possibilities, either the $4$
supertransformations lying on the Fano plane are complementary to a line or $3$ of them belong 
to one of the lines. In this case to get the minimal ${\cal N}=4$ decomposition we have to pick them together with $Q_8$.   \par
Well, what all this has to do with chirality and quaternionic chirality? The fact is that, in all cases,
no matter which decomposition into two minimal ${\cal N}=4$ supermultiplets is taken, one
always ends up with two ${\cal N}=4$ root supermultiplets of opposite chirality.
The chirality, which is irrelevant when a single supermultiplet is concerned, suddenly turns out
to be crucial when several copies of them are considered. The ${\cal N}=5,6,7$ supermultiplets are essentially
non-chiral because in their ${\cal N}=4$ minimal decomposition two multiplets, a quaternionic chiral and a
quaternionic antichiral one, are produced. The same is true for the ${\cal N}=8$
root supermultiplet. It is quaternionic non-chiral in the sense here specified. For ${\cal N}=8$ on the other
hand, another notion of chirality, the octonionic chirality, can be introduced (no such notion makes sense for ${\cal N}=5,6,7$). The ${\cal N}=8$ root supermultiplet is quaternionically
non-chiral and octonionically chiral.\par
This result admits the following reformulation: given two minimal ${\cal N}=4$ supermultiplets,
they can be combined into a single minimal supermultiplet for ${\cal N}=5,6,7$ or $8$ if and only if the two supermultiplets possess opposite chirality. Indeed, if one tries to link together
with an extra supersymmetry transformation two supermultiplets of the same chirality, one
ends up in a contradiction.\par
Two chiral supermultiplets are isomorphic (via the $Z_2$ isomorphism) with two antichiral
supermultiplets, while they are not isomorphic with a chiral and an antichiral supermultiplet,
due to the fact that one cannot flip the chirality of the first multiplet without flipping the
chirality of the second multiplet. The reason is that the $Z_2$ isomorphism requires
a relabeling of the supertransformations, as discussed in the previous section, not just a 
transformation of the fields entering the supermultiplets (as it is the case for ${\cal N}=3$).
\par
Given a certain number $n$ of ${\cal N}=4$ $(4,4)$ root supermultiplets, inequivalent chiralities are discriminated by
the modulus (to make it $Z_2$-invariant) $m$ given by
\bea
m &=& |\sum_is_i|= |n_+-n_-|,
\eea
where $n=n_++n_-$, with $n_\pm$ denoting the number of chiral (antichiral) supermultiplets.\par
For $n=1$ we have that $m=1$. The non-chiral case ($m=0$) is only possible for $n$ even.\par
It is certainly possible and perhaps even convenient to think of $m$ as an energy. In this interpretation the non-chiral state corresponds to the vacuum. For odd $n$, the vacuum energy
is positive. The vacuum is doubly degenerated and spontaneously broken. For the single
($n=1$) supermultiplet the two (equivalent) choices of the chirality ($\pm 1$) can be interpreted as a spontaneous breaking of the $Z_2$ symmetry.\par
For a collection of $n$ root supermultiplets of ${\cal N}=8$, the previous steps are repeated
in terms of the notion of ``octonionic chirality" (each ${\cal N}=8$ supermultiplet is non-chiral
for what concerns the quaternionic chirality).

\section{Weyl on the Leibniz-Clarke debate on the nature of space}

The problem of understanding the nature of chirality for quaternionic (anti)chiral ${\cal N}=4$
root supermultiplets and octonionic (anti)chiral ${\cal N}=8$ root supermultiplets is similar
to the problem of understanding the nature of parity (mirror symmetry) of the Euclidean space.
A nice framework was provided by Weyl in his popular book on Symmetry \cite{weyl}. \par
The famous Clarke-Leibniz debate concerning the nature of the space (either absolute, thesis defended by the Newtonians, Clarke was one of them) or relative (thesis defended by Leibniz who anticipated some of the arguments later used by Mach) is well-known. The Clarke-Leibniz debate was expressed in the metaphysical and theological language of the time. Weyl, in his book, reformulates the position of Leibniz (and also Kant) by expressing the relative nature of the space for the specific $Z_2$ parity transformation associated with the mirror symmetry. Weyl,
in his argument, mimicks the theological framework used by Clarke and Leibniz. \par
The Weyl argument goes as follows. Let us suppose that God at the beginning creates out of nothing, in the empty space, a hand. We have no way to say whether this hand is left or right
(in the empty space the hand is so good as its mirror image). Now, let us suppose that after
creating the first hand, God creates a second hand. It is only after this second hand has been created that the notion of right or left has been introduced. The second hand can be aligned
with the first one or be of opposite type. Right-handedness or left-handedness is a relative notion
based on the referential provided by the first hand.\par
The argument of Weyl clearly applies to the notion of quaternionic (or octonionic) chirality
for ${\cal N}=4$ (${\cal N}=8$) supermultiplets. In the case of supermultiplets on the other
hand we can still go further. Let us focus on the quaternionic chirality of the ${\cal N}=4$
supermultiplets. We can go on with theological speculation. Let us suppose now that, after the
second hand has been created (so that we have now two hands floating in the empty space) God performs a third act of creation, creating a handless body. This handless body, floating in the empty space, try to grasp and attach the two floating hands to its handless arms. He/she can only do that if the two hands are of opposite chirality. He/she is unable to do that if they come
with the same chirality. Needless to say, the handless body can stay for the full ${\cal N}=8$
supersymmetry which can be obtained by linking together two minimal ${\cal N}=4$ root supermultiplets. With this baroque image of handless bodies desperately seeking floating hands
in the empty space, we can leave the Clarke-Leibniz-Weyl theological speculations.

\section{Conclusions}

In this paper I clarified the issue of the chirality associated with ${\cal N}=4$ and ${\cal N}=8$
supermultiplets.  I pointed out that there are two notions of chirality which can be introduced:
quaternionic chirality which applies to ${\cal N}=4,5,6,7,8$ minimal supermultiplets and octonionic chirality which applies to ${\cal N}\geq 8 $ minimal supermultiplets. For ${\cal N}>4$
the minimal supermultiplets are quaternionically nonchiral (every decomposition into two minimal ${\cal N}=4$ minimal supermultiplets produces two supermultiplets of opposite chirality). The same is true (concerning octonionic chirality) for the minimal supermultiplets with ${\cal N}>8$. The notion of quaternionic chirality applies to the ${\cal N}=4$ minimal supermultiplets, while the notion of octonionic chirality
to the minimal ${\cal N}=8$ supermultiplets. \par
In both cases a {\em single} chiral supermultiplet is isomorphic to its antichiral counterpart
via a $Z_2$ transformation. Chirality cannot be detected if we deal with a single ${\cal N}=4$
(${\cal N}=8$) supermultiplet. \par
On the other hand, the chirality issue becomes important when we are dealing with reducible
${\cal N}=4$ (${\cal N}=8$) supermultiplets given by several copies of chiral and antichiral
supermultiplets. 
Let us take the example of the reducible ${\cal N}=4$ representation given by two separate
root supermultiplets (the total field content is $(8,8)= 2\times (4,4)$). There are two inequivalent
such reducible representations ($(8,8)_{red., ch.}$ and $(8,8)_{red., nc.}$). One is chiral
($(8,8)_{red., ch.}$); it is given by two ${\cal N}=4$ root supermultiplets of the same chirality.
The other one ($(8,8)_{red., nc.}$) is nonchiral and given by two ${\cal N}=4$ root supermultiplets of opposite chirality. Only $(8,8)_{red., nc.}$ can be ``promoted" to a minimal
irreducible representation for ${\cal N}=5,6,7,8$ by inserting extra supertransformations
linking its two ${\cal N}=4$ component supermultiplets.\par
The overall chirality of $n$ ${\cal N}=4$ supermultiplets is important when constructing
${\cal N}=4$ off-shell invariant actions (sigma-models) for this large set of $n$ supermultiplets.\par
The notion of quaternionic chirality for root supermultiplets gets extended to the remaining 
linear supermultiplets which are obtained by dressing (their chirality is encoded in the chirality
of their associated root supermultiplets) . It is also extended to the two nonlinear ${\cal N}=4$
supermultiplets ($(3,4,1)_{nl}$ and $(2,4,2)_{nl}$) which are also produced, see \cite{fkt}, from  
the ${\cal N}=4$ root supermultiplet.  The chirality of the originating root supermultiplet
has to be taken into account.

\appendix

For completeness we report the definitions, applied to the cases used in the text, of the properties
characterizing the linear representations of the one-dimensional ${\cal N}$-Extended Superalgebra.
In particular the notions of {\em mass-dimension}, {\em field content}, {\em dressing transformation},
{\em connectivity symbol}, {\em dual supermultiplet} and so on, as well as the association of linear
supersymmetry transformations with graphs, will be reviewed following \cite{{pt},{krt},{fg},{kt},{kt2},{gkt}}.
The Reader can consult these papers for broader definitions and more detailed discussions. \par
{\em Mass-dimension}:\par
A grading, the mass-dimension $d$, can be assigned to any field entering a linear representation
(the hamiltonian $H$, proportional to the time-derivative operator $\partial\equiv \frac{d}{dt}$,
has a mass-dimension
$1$). Bosonic (fermionic) fields have integer (respectively, half-integer) mass-dimension. \par
{\em Field content:}\par
Each finite
linear representation is characterized by its ``field content", i.e. the set of integers $(n_1,n_2,\ldots , n_l)$
specifying the number $n_i$ of fields of mass-dimension
$d_i$ ($d_i = d_1 + \frac{i-1}{2}$, with $d_1$ an arbitrary constant) entering the representation.
Physically, the $n_l$ fields of highest dimension are the auxiliary fields which transform as a time-derivative
under any supersymmetry generator. The maximal value $l$ (corresponding to the maximal dimensionality
$d_l$) is defined to be the {\em length} of the representation (a root representation has length $l=2$).
Either $n_1, n_3,\ldots$ correspond to the bosonic fields (therefore $n_2, n_4, \ldots$ specify the fermionic
fields)
or viceversa. \\ In both cases the equality $n_1+n_3+\ldots =n_2+n_4+\ldots = n$ is guaranteed.\par
\par
{\em Dressing transformation:}\par
Higher-length supermultiplets are obtained by applying a dressing transformation to the
length-$2$ root supermultiplet. The root supermultiplet is specified by the ${\cal N}$ supersymmetry
operators ${\widehat Q}_i$ ($i=1,\ldots ,{\cal N})$, expressed in matrix form as
\begin{eqnarray}\label{hatq}
{\widehat Q}_j = \frac{1}{\sqrt{2}}\left(
\begin{array}{cc}
0&\gamma_j\\
-\gamma_j\cdot H&0
\end{array}
\right), &&
{\widehat Q}_{\cal N} =\frac{1}{\sqrt{2}} \left(
\begin{array}{cc}
0&{\bf 1}_n\\
{\bf 1}_n\cdot H&0
\end{array}
\right),
\end{eqnarray}
where  the $\gamma_j$ matrices ($j=1,\ldots , {\cal N}-1$) satisfy the Euclidean Clifford algebra
\begin{eqnarray}
\{\gamma_i,\gamma_j\}&=& -2\delta_{ij}{\bf 1}_n.
\end{eqnarray}
The length-$3$ supermultiplets are specified by the ${\cal N}$ operators $Q_i$, given by the dressing transformation
\begin{eqnarray}\label{dressingtransformation}
Q_i &=& D{\widehat Q}_i D^{-1},
\end{eqnarray}
where $D$ is a diagonal dressing matrix such that
\begin{eqnarray}\label{dressingmatrix}
{D} &=& \left(
\begin{array}{cc}
{\widetilde{D}}&0\\
0&{\bf 1}_n
\end{array}
\right),
\end{eqnarray}
with ${\widetilde D}$ an $n\times n$ diagonal matrix whose diagonal entries are either $1$ or the derivative
operator $\partial$. \par
{\em Association with graphs:}\par
The association between linear supersymmetry transformations and ${\cal N}$-colored oriented graphs goes as
follows. The fields (bosonic and fermionic) entering a representation
are expressed as vertices. They can be accommodated into an $X-Y$ plane. The $Y$ coordinate
can be chosen to
correspond to the mass-dimension $d$ of the fields. Conventionally, the lowest dimensional fields
can be
associated to vertices lying on the $X$ axis. The higher dimensional fields have positive, integer or
half-integer values of $Y$.
A colored edge links two vertices which are connected by a supersymmetry transformation. Each one of the ${\cal N}$ $Q_i$
supersymmetry generators is associated to a given color. The edges are oriented. The orientation reflects the sign
(positive or negative) of the corresponding supersymmetry transformation connecting the two vertices. Instead
of using
arrows, alternatively, solid or dashed lines can be associated, respectively, to positive or negative signs.
No colored line is drawn for supersymmetry transformations connecting a field with the time-derivative of a
lower
dimensional field. This is in particular true for the auxiliary fields (the fields of highest dimension in the
representation) which are necessarily mapped, under supersymmetry transformations, in the time-derivative
of lower-dimensional fields.
\par
Each irreducible supersymmetry transformation can be presented (the identification is not unique) through
an oriented
${\cal N}$-colored graph with $2n$ vertices. The graph is such that precisely ${\cal N}$ edges, one for each
color, are linked to any given vertex which represents either a $0$-mass dimension or a $\frac{1}{2}$-mass
dimension field.  An unoriented ``color-blind" graph can be associated to the initial graph by disregarding
the orientation of the edges
and their colors (all edges are painted in black). \par
{\em Connectivity symbol:}\par
A characterization of length $l=3$ color-blind, unoriented graphs can be expressed through the connectivity
symbol
$\psi_g$, defined as follows
\begin{eqnarray}
\psi_g &=& ({m_1})_{s_1} +({m_2})_{s_2}+\ldots +({m_Z})_{s_Z}.
\end{eqnarray}
The $\psi_g$ symbol encodes the information on the partition of the $n$  $\frac{1}{2}$-mass dimension fields
(vertices)
into the sets of $m_z$ vertices ($z=1,\ldots, Z$) with $s_z$ edges connecting them to the $n-k$ $1$-mass
dimension auxiliary fields.
We have
\begin{eqnarray}
m_1+m_2+\ldots +m_Z &=& n,
\end{eqnarray}
while $
s_z\neq s_{z'}$ for $  z\neq z'$.\par
{\em Dual supermultiplet:}\par
A dual supermultiplet is obtained by mirror-reversing, upside-down, the graph associated
to the original supermultiplet.

\section*{Acknowledgments}
% Replace next line with your Acknowledgements:
This work has been supported by CNPq.

% References go below:

\end{document}